\documentclass[twocolumn,final,aps,prl,showpacs,superscriptaddress,amsmath,amssymb,amsfonts,floatfix]{revtex4-1}

\usepackage{graphicx}
\usepackage{multirow}
\usepackage{epsfig}
\usepackage{url}
\usepackage{color}

\usepackage[colorlinks,breaklinks,bookmarks=true,citecolor=blue,linkcolor=blue,urlcolor=blue]{hyperref}
\usepackage{color}
\usepackage{tabularx}
\usepackage{sidecap}
\usepackage{soul}

\newcommand{\code}[1]{\texttt{#1}}

\newcommand{\plane}{Ba$_2$CuO$_3$}
\newcommand{\original}{Ba$_2$CuO$_4$}
\newcommand{\bilayer}{Ba$_2$CuO$_{3.25}$}
\newcommand{\experiment}{Ba$_2$CuO$_{3.2}$}
\newcommand{\LCO}{La$_2$CuO$_4$}

\newcommand{\dbc}{$d_{b^2-c^2}$}

\usepackage{tikz}
\usetikzlibrary{calc}
\usetikzlibrary{arrows}
\usetikzlibrary{patterns}
\usetikzlibrary{decorations.pathmorphing}
\usetikzlibrary{decorations.pathreplacing}
\usetikzlibrary{decorations.markings}
\usetikzlibrary{shapes.geometric}
\usetikzlibrary{positioning}

    \tikzset{middlearrow/.style={
                decoration={markings,
                            mark= at position 0.65 with {\arrow{#1}},
                                    },
                                            postaction={decorate}
                                                }
                                                }

\begin{document}

\author{Paul Worm}
\affiliation{Institute of Solid State Physics, TU Wien, 1040 Vienna, Austria}

\author{Motoharu Kitatani }
\affiliation{RIKEN Center for Emergent Matter Science (CEMS), Wako, Saitama, Japan}

\author{Jan M. Tomczak}
\affiliation{Institute of Solid State Physics, TU Wien, 1040 Vienna, Austria}

\author{Liang Si}
\email{liang.si@ifp.tuwien.ac.at}
\affiliation{Institute of Solid State Physics, TU Wien, 1040 Vienna, Austria}

\author{Karsten Held}
\affiliation{Institute of Solid State Physics, TU Wien, 1040 Vienna, Austria}

\title{Hidden, one-dimensional, strongly nested, and almost half-filled Fermi surface  in Ba$_{2}$CuO$_{3+y}$ superconductors}

\begin{abstract}

All previous cuprate superconductors display a set of common features: (i) vicinity to a Cu 3$d^{9}$ configuration; (ii) separated CuO$_2$ planes; (iii) superconductivity for doping $\delta \sim$ 0.1$-$0.3. Recently [PNAS {\bf 24}, 12156 (2019)] challenged this picture by discovering ``highly overdoped'' superconducting Ba$_2$CuO$_{3+y}$. Using density-functional theory + dynamical mean-field theory, we reveal a bilayer structure of Ba$_2$CuO$_{3.2}$ of alternating quasi 2D and quasi 1D character. Correlations tune an inter-layer self-doping leading to an almost half-filled, strongly nested  quasi 1D $d_{b^2-c^2}$ band, which  is prone to strong antiferromagnetic fluctuations, 
possibly at the origin of superconductivity in Ba$_2$CuO$_{3+y}$.

\end{abstract}

\maketitle

\textit{Introduction} -----
Even 35 years after the discovery of high-temperature superconductivity in cuprates \cite{Bednorz1986}, the   pairing mechanism  remains highly controversial.
In this respect, the recently discovered cuprate \experiment~\cite{Li12156}
is supremely exciting as it puts into question common wisdom for cuprate high-temperature superconductivity. The oxygen reduction from \original\ to polycrystalline Ba$_2$CuO$_{3+y}$ was achieved by synthesizing the samples at extremely high pressure ($\sim18\,$GPa) and high temperature ($\sim1000^{\circ}\,$C). Unusual is, first of all, the hole concentration in the superconducting  $y\sim0.2$ phase which has $\delta=2y=0.4$ holes with respect to a Cu 3$d^9$ electronic configuration. This is twice as many holes as in other superconducting cuprates. Despite this unusual doping the critical temperature $T_c=70\,$K is high \cite{Li12156}.

Second, a \LCO-type structure with space group I4/mm was suggested by the authors of \cite{Li12156}, with a compressed oxygen octahedron, contrary to an elongated one.  
This compression also pushes the $3z^2-r^2$ orbital, that is fully occupied in other cuprates, up in energy. More recent experiments \cite{Fumagalli2021} report, however, x-ray absorption (XAS) and resonant inelastic x-ray scattering (RIXS) data, incompatible with the \LCO~structure. These results \cite{Fumagalli2021} require two inequivalent Cu sites, proposing a bilayer structure. Unfortunately, single crystals have not yet been synthesized. This leaves quite an uncertainty even for the crystal structure and different ones have been suggested \cite{Li12156,Fumagalli2021,Liu2019,Wang2020,Jiang2021,Li2020,Kimihiro2020}.

\begin{figure}[tb]
\centering
\includegraphics[width=0.5\textwidth]{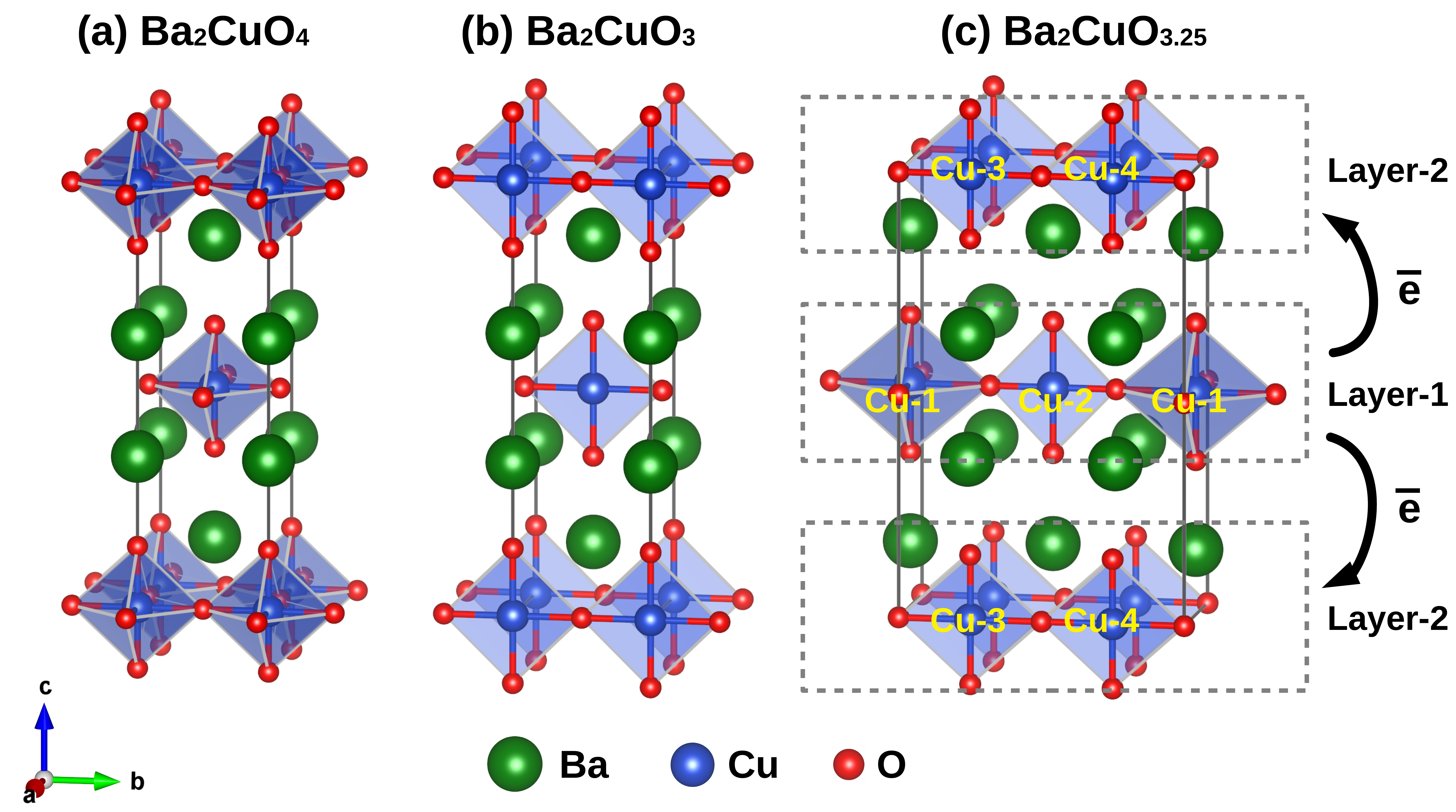}
\caption{Crystal structure of (a) the parent compound \original\ and (b) the ideal, fully reduced \plane, where the planar oxygen atoms in $a$ direction are vacant. (c) Energetically favorable crystal structure for \bilayer~\cite{Liu2019} close to the superconducting doping $y=0.2$. Note (c) forms a bilayer structure, layer-2 has as (b) the in-plane oxygens in the $a$ direction removed, in layer-1 every second one is removed. This leads in layer-2 to 1D Cu-3---O---Cu-4---O--$\cdots$ chains. Arrows indicate the inter-layer charge transfer that is driven by electronic correlations.}
\label{fig:Structure}
\end{figure}

Guided by the experimental results and previous density functional theory (DFT)  calculations \cite{Liu2019}, we investigate the electronic structure of the three crystal structures of Fig.~\ref{fig:Structure}: (a) the parent compound \original, (b) the fully reduced material \plane, and (c) a bilayer-structure \bilayer. The last structure has an  oxygen deficiency (excess) of $0.75$ ($y=0.25$) compared to the structure of panel Fig.~\ref{fig:Structure}(a) ((b)). It is close to  $y = 0.2$ but can be realized in a smaller 2$\times$2$\times$1 unit cell by removing three oxygens (adding one oxygen). To find the ground state structure of Ba$_2$CuO$_{3+y}$ near $y$=0.2, we consider all variations proposed in previous studies  \cite{Liu2019,Kimihiro2020} and find the bilayer structure of Fig.~\ref{fig:Structure}(c) to be the energetically most favorable among all possible 2×2×1 super cells with four Cu sites.


%
The primitive cell  Fig.~\ref{fig:Structure}(c) can be obtained from \plane\ [Fig.~\ref{fig:Structure}(b)] by inserting one oxygen into the empty spaces in the ``layer-1'' CuO$_2$ planes at the Cu-1 sites. This results in a Ba$_8$Cu$_4$O$_{13}$ supercell that contains four Cu sites, resulting in the chemical formula \bilayer. It is composed of two different layers: in layer-1 we have Cu-1 sites with a six-fold octahedral CuO$_6$ coordination, and Cu-2 sites with planar CuO$_4$ squares. In layer-2 both  Cu-3$\&$4 sites are equivalent and the same as in \plane\ [Fig.~\ref{fig:Structure}(b)] with planar CuO$_4$ squares. They form 1-dimensional (1D) CuO chains in the $b$ direction. If we compare to the parent compound \original, oxygen reduction has removed the planar O in the $a$ direction for Cu-2, Cu-3, and Cu-4 sites. Please note that removing parts of the oxygen atoms from the CuO$_2$ planes and forming the 1D CuO chains, will result in an orthorhombic distortion of Ba$_2$CuO$_{3.2}$, if all CuO chains point in the same direction, as discussed previously for Ba$_2$CuO$_3$ \cite{yamamoto2000new}. However, the synthesizing process of oxygen reduced Ba$_2$CuO$_{3+y}$ under high temperature and high pressure might stabilize an undistorted crystal where the CuO chains in different layers point in alternating directions. The structure is further stabilized by the fact that every other layer still contains $a$-$b$ symmetric CuO$_6$ octahedra.

In this paper, we present DFT and DFT+dynamical mean-field theory (DMFT) \cite{RevModPhys.68.13,kotliar2004strongly,held2007electronic} calculations for all three crystal structures of Fig.~\ref{fig:Structure} as well as for the superconducting  \experiment.
To obtain the hole doping of the latter, we employ a rigid (bandstructure)  doping. We find that the physics is completely different for the three structures: \original\ is a two-orbital system, while \plane\ a one-orbital 1D system. The \bilayer\ supercell inherits aspects of both parent compounds in its two inequivalent layers. Crucially, correlations induce a charge transfer so that  layer-2 of \bilayer\ is doped close to half-filling and prone to strong antiferromagnetic spin-fluctuations.

\textit{Methods} -----
DFT-level computations are performed by  \code{WIEN2K} \cite{blaha2001wien2k,schwarz02} using the Perdew-Burke-Ernzerhof \cite{Perdew2008} version of the generalized gradient approximation (GGA-PBE).
The structural parameters of ideal Ba$_2$CuO$_4$ and Ba$_2$CuO$_3$ are adopted from Refs.\
\cite{Li12156,khoroshilov1994crystal}, and the crystal structure of the bilayer \bilayer~phase was optimized within DFT-PBE. As an input for the DMFT calculations a low-energy effective Hamiltonian is generated by projecting the \code{WIEN2K} bands around the Fermi level onto Wannier functions  \cite{PhysRev.52.191,RevModPhys.84.1419} using \code{WIEN2WANNIER}  \cite{mostofi2008wannier90,kunevs2010wien2wannier}. These are supplemented by a local Kanamori interaction and the fully localized limit as double counting \cite{Anisimov1991}. The constrained random phase approximation (cRPA) \cite{PhysRevB.77.085122} yields:  average intraorbital Hubbard interaction $U = 2.6$\,eV, Hund's exchange $J = 0.3$\,eV; the interorbital interaction follows as $U' = U - 2J$. For our calculations we used slightly enhanced Hubbard interactions $U = 3.0$\,eV to mimic the disregarded frequency dependence of $U$\hspace{-0.5mm}, along the lines of \cite{Kitatani2020} and many other publications. We solve the resulting many-body Hamiltonian at room temperature (300\,K) within DMFT employing a continuous-time quantum Monte Carlo solver in the hybridization expansions \cite{RevModPhys.83.349} with \code{w2dynamics} \cite{PhysRevB.86.155158,wallerberger2019w2dynamics}. Real-frequency spectra are obtained with the \code{ana\_cont} code  \cite{kaufmann2021anacont} via analytic continuation using the maximum entropy method (MaxEnt).

\begin{figure}[t]
\centering
\includegraphics[width=0.49\textwidth]{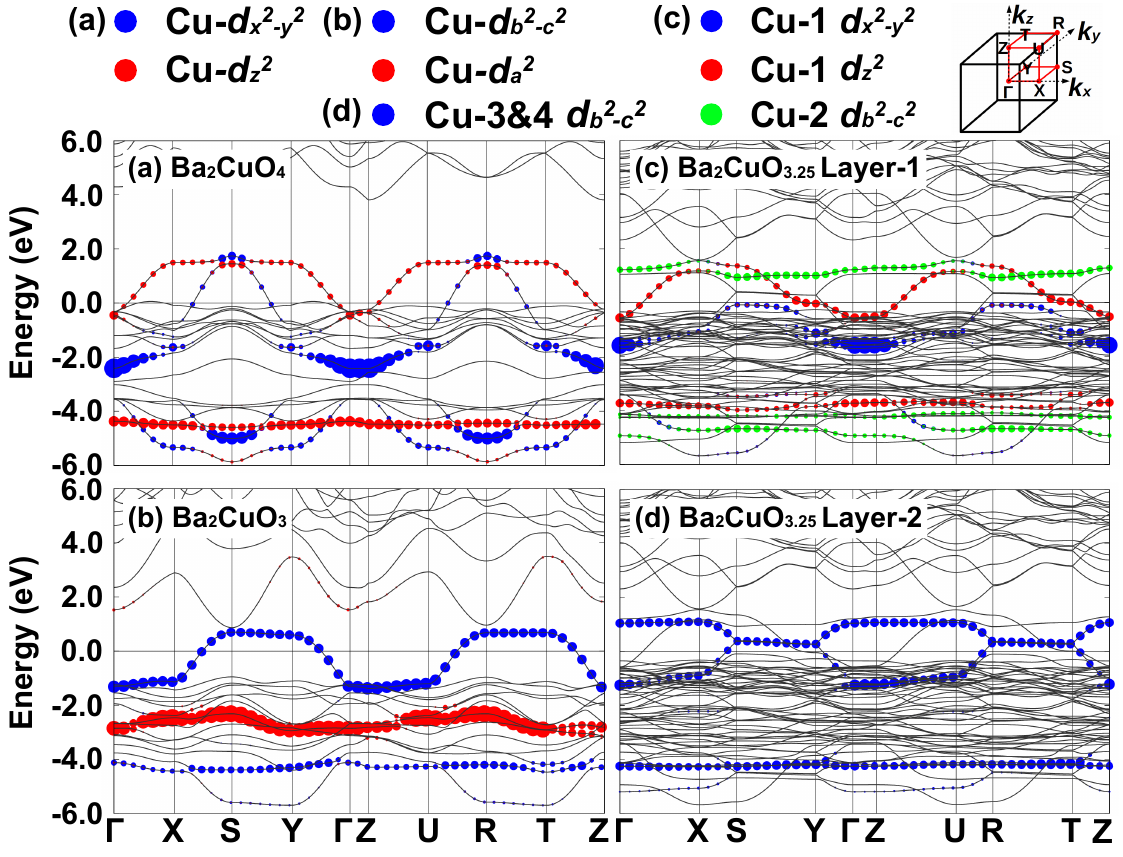}
\caption{DFT bandstructure and orbital character for (a) \original, (b) \plane, (c) \bilayer~layer-1  and (d) \bilayer~layer-2  along a high symmetry path through the Brillouin zone (see top right).}
\label{fig:DFT_Band}
\end{figure}


\textit{DFT electronic structure} ----- 
Let us first review the DFT electronic structure for the three different crystal structures. For \original\ the Fermi surface (FS)  in Fig.~\ref{fig:DFT_Band}(a) is composed of two Cu-$d$ bands of $d_{x^2-y^2}$ and $d_{z^2}$ orbital character, consistent with previous results \cite{Maier_2019,Li12156} and hinting toward multi-orbital physics. Instead,
\plane\ in Fig.~\ref{fig:DFT_Band}(b) hosts only one Fermi surface sheet of $d_{b^2-c^2}$ orbital character \cite{Liu2019} (this orbital is like a $x^2-y^2$ orbital only in the $bc$ plane as the missing oxygen in the $a$ direction dictates the local symmetry). Because the Cu layers are well separated in the $c$ direction, this leads to a  quasi 1D character of the bandstructure in the $b$ direction. Superconductivity was, however, observed in neither of these two parent compounds, but at an oxygen concentration $y=0.2$ for Ba$_2$CuO$_{3.2}$. A rigid band shift of the \plane\ or \original\ bandstructure to this $y=0.2$ doping results in two profoundly different FSs (see Fig.~\ref{fig:Fig5_DMFT_Fermi_Surface} below), electronic structures and even orbital occupations. This naturally prompts the question: Does the FS of Ba$_2$CuO$_{3.2}$  show \plane\ or \original\ character? To address this question, we perform a supercell calculation, which allows for non-uniform oxygen reduction \cite{Liu2019}. We identify the most promising structure for \bilayer\ (which is reasonably close to one of the experimental oxygen concentrations) to be the one in Fig.~\ref{fig:Structure}(c), based on DFT total energy. 
This structure has two inequivalent layers, each of which is similar to the two parent compounds: 
The Cu-1 sites in layer-1 have the same  local octahedra as in \original, and its low-energy excitations  in Fig.~\ref{fig:DFT_Band}(c) are hence similarly described by a  Cu-$d_{x^2-y^2}$ and a Cu-$d_{z^2}$ orbital. The Cu-2 site in layer-1 has an oxygen removed and hosts a $d_{b^2-c^2}$ band, which does not cross the Fermi level in Fig.~\ref{fig:DFT_Band}(c). The  Cu-3$\&$4 sites in layer-2, on the other hand, display the same local surroundings as \plane\ and  their low-energy electronic structure as in Fig.~\ref{fig:DFT_Band}(d) is described by a single \dbc\ orbital \cite{PhysRevB.44.6011,PhysRevB.57.10287}. This band is similar to Fig.~\ref{fig:DFT_Band}(b), see also hopping elements in Table~\ref{table:hopping}.

\begin{table}
\caption{Major hopping elements in meV for the \dbc~ orbital in \plane~ and layer-1 of \bilayer. \label{table:hopping}}
\begin{tabular}{ cccccc }
\hline
\hline
$ \quad ~ \quad$ Structure $\quad \quad $ & $ \quad t_a \quad $   & $ \quad t''_a  \quad$ & $ \quad t_b \quad$ & $ \quad t''_b  \quad$ & $ \quad t'_{ab}  \quad$\\
\hline
\plane    & -18.5 & -1.3 & -470.2 & -84.6 & -6.8 \\
\bilayer  & -25.8  & 1.4 & -518.1 & -89.4 & -11.9 \\
\hline
\end{tabular}
\label{Tab:HoppingElements}
\end{table}

\begin{figure}[t]
\centering
\includegraphics[width=0.49\textwidth]{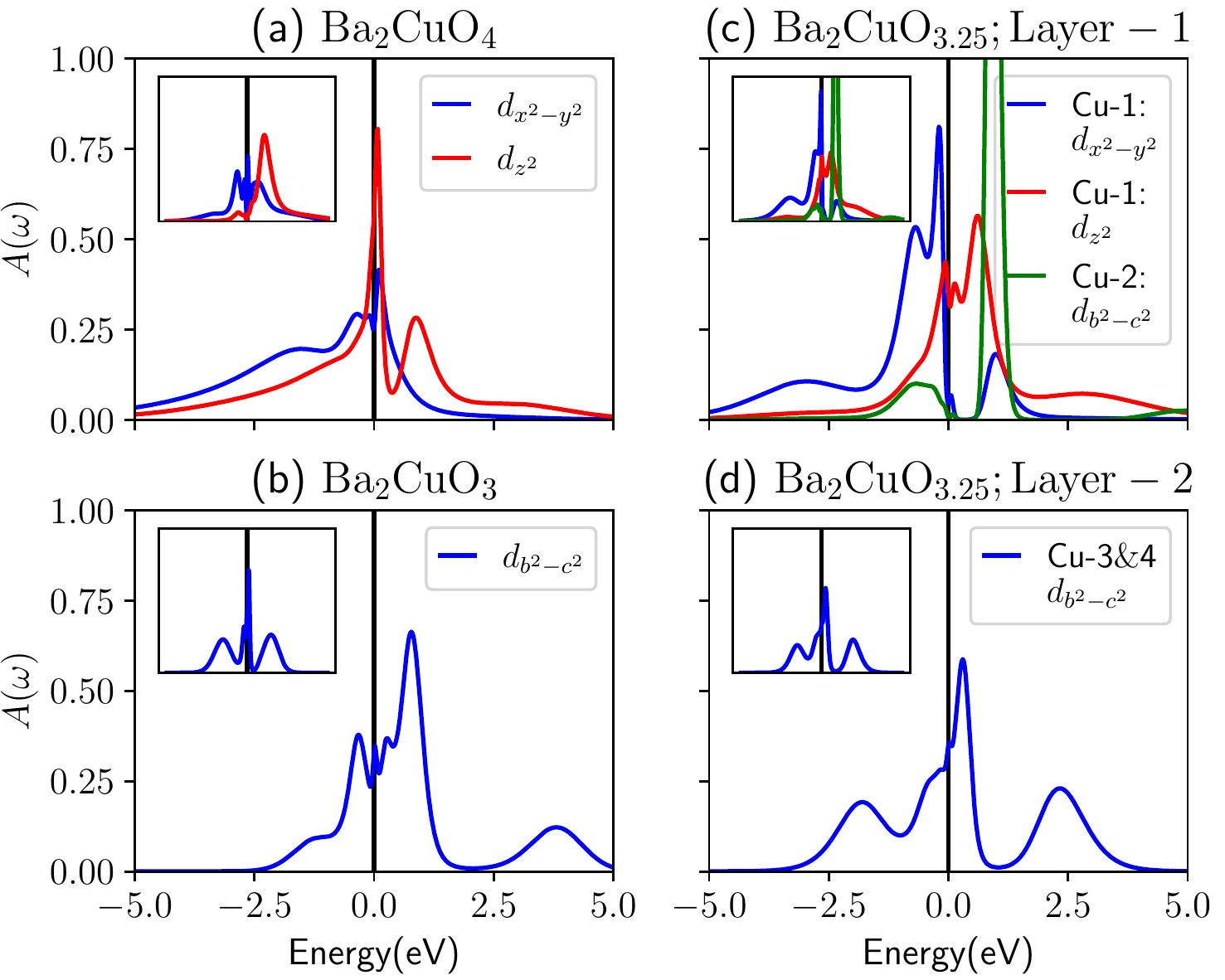}
\caption{DMFT spectral function $A(\omega)$ for the Wannier-projected low-energy bands of Fig.~\ref{fig:DFT_Band} three crystal structures: (a) \plane, (b) \original, (c) \bilayer\ layer-1 and (d) \bilayer\ layer-2. The inset shows results for the stoichiometric parent compound,  the main panel corresponds to the hole doping of \experiment. Insets have the same scale as main panels.}
\label{fig:DMFT_Spectrum}
\end{figure}

\textit{DFT+DMFT electronic structure} -----
Since cuprates are known for strong correlation effects, we expect significant corrections to the DFT results. To address these we perform DFT+DMFT calculations in the paramagnetic phase at room temperature (300\,K). The DMFT momentum-integrated spectral function $A(\omega)$ is displayed in Fig.~\ref{fig:DMFT_Spectrum}. We show the spectrum both for the stoichiometric parent compound (inset) and an adjusted particle number  (main panel) to reach an oxygen concentration of $y=0.2$.  The arguably simplest system is undoped \plane~[Fig.~\ref{fig:DMFT_Spectrum}(b) inset] which  shows a single low-energy orbital with the typical three-peak spectrum of correlated electron systems. Besides a lower and an upper Hubbard band, there is a central quasiparticle peak with mass enhancement $m^*/m \equiv 1/Z \sim 3.85$. Please note that we are considering the paramagnetic solution, even though without doping \plane~has a strong tendency to antiferromagnetism  \cite{Liu2019}. Upon doping, see Fig.~\ref{fig:DMFT_Spectrum}(b),  correlation effects become much weaker as evidenced by a reduced mass enhancement of $m^*/m \sim 1.45$.
The other parent compound, \original~in Fig.~\ref{fig:DMFT_Spectrum}(a), hosts two orbitals: a moderately correlated $d_{x^2-y^2}$ orbital close to half-filling ((spin-summed) occupation $n \sim 0.87$, $m^*/m\sim 1.39$) and a weakly correlated $d_{z^2}$ orbital ($n \sim 0.13$, $m^*/m\sim 1.15$). A metallic behavior at this doping was also reported previously in \cite{PhysRevB.103.214510}. Removing $1-y=0.8$ oxygen per formula, electrons dope both orbitals, slightly increase the mass enhancement in the $d_{x^2-y^2}$ orbital ($n \sim 1.56$, $m^*/m\sim 1.47$) while dramatically boosting it for the $d_{z^2}$ orbital ($n \sim 1.04$, $m^*/m\sim3.12$).

We now assess how these trends survive in the structurally akin Cu-O planes of the bilayer compound \bilayer. Let us thus turn to Fig.~\ref{fig:DMFT_Spectrum}(c) and (d) which show layer-1 and layer-2 spectra of the bilayer structure \bilayer, respectively. Here, correlations play a crucial role: they drive a charge transfer from layer-1 to layer-2. As a consequence the occupation of the \dbc\ orbital in layer-2 is increased towards half-filling ($n \sim$ $0.91$). This leads to a strongly correlated spectrum in Fig.~\ref{fig:DMFT_Spectrum}(d) with $m^*/m\sim 2.40$. For \plane, we would have a similar $m^*/m\sim 2.63$, if the occupation is fixed at 0.9 electrons per \dbc\ orbital. Hence, we conclude that the strongly correlated, half-filled, single \dbc\ band physics is preserved in the \experiment\ structure. As superconductivity is known to be extremely sensitive to doping \cite{Rybicki2016,Qin2021}, let us now discuss this important charge transfer in more detail.

\textit{Discussion: correlation-induced charge transfer} -----
One common ingredient for cuprate and recently observed nickelate superconductors \cite{li2019superconductivity,Kitatani2020} has been the CuO$_2$ or NiO$_2$ plane, whose low-energy physics is dominated by a $d_{x^2-y^2}$ orbital close to half-filling \cite{Bednorz1986, anderson1987resonating}. Neither \plane~nor \original~fits into this CuO$_2$ plane category. The former hosts an overdoped, quasi 1D $d_{b^2-c^2}$ band, while the latter displays  two-orbital, $d_{x^2-y^2}$ and $d_{z^2}$ physics. Both compounds have been studied theoretically and several mechanisms for the superconductivity have already been proposed \cite{Maier_2019, Kimihiro2020, Jiang2021, gao2020nodeless, klett2021high, PhysRevB.103.214510,PhysRevB.103.144511}. Let us stress, that a rigid bandstructure doping from either parent compound will always be plagued by the ambiguity of which structure is realized for the experimental \experiment\ compound, especially since both parent compounds are far away from the superconducting  oxygen content. This problem is resolved by turning to a supercell calculation. Here, the stabilization of a bilayer structure is crucial. However, a naive electron count for the three inequivalent Cu sites of Fig.~\ref{fig:Structure}(c) would be Cu-1: $d^7$, Cu-2: $d^9$ and Cu-3$\&$4: $d^9$ when considering the local CuO$_6$ and CuO$_4$ configurations. A charge transfer between Cu-1 and Cu-2 can be expected, as they are located in the same layer and connected by oxygen. Somewhat less straight-forward, but arguably more interesting is the inter-layer charge transfer.  In DFT, for \bilayer, about $\sim 0.44$ electrons will relocate from layer-2 into layer-1, which results in an occupation of $\sim 0.78$ for the Cu-3$\&$4 $d_{b^2-c^2}$ orbitals.

Local DMFT correlations of counteract this charge transfer. They favor an even distribution of electrons among the orbitals, see Table~\ref{Tab1:Occupation}. Specifically, $\sim$ 0.12 electrons relocate back to the  Cu-3$\&$4 in layer-2, leading to 0.84 electron per $d_{b^2-c^2}$. As the oxygen content of \bilayer\ is already close to the experimental compound Ba$2$CuO$_{3.2}$, a rigid  doping to \experiment\ is more justified here, and the occupation of the various orbitals are also listed in Table~\ref{Tab1:Occupation}. Of particular interest are the Cu-3$\&$4 orbitals in layer-2, which are now unexpectedly close to typical doping levels of common quasi 2D cuprate superconductors.

\begin{table}[]
\caption{DFT and DMFT electron occupations for the Cu sites of \bilayer\ and its rigid doping to \experiment.}
\begin{tabular}{cc|cc|cc}
\hline
\hline
 Site &  Orbital   & \multicolumn{2}{c|}{\quad Ba$_2$CuO$_{3.25}$ \quad}  &      \multicolumn{2}{c}{\quad Ba$_2$CuO$_{3.20}$ \quad} \\ \hline
   &              &    \quad   DFT  \quad  &  \quad DMFT \quad        &  \quad DFT \quad  & \quad  DMFT  \quad                 \\ \hline
Cu-1 & $d_{x^2-y^2}$           & 1.70      & 1.65      & 1.75 & 1.73                 \\ 
Cu-1 & $d_{z^2}$               & 0.50      & 0.50      & 0.64 & 0.63                 \\ 
Cu-1 & $d_{x^2-y^2}$+$d_{z^2}$ & 2.20      & 2.15      & 2.40 & 2.37                 \\ 
Cu-2 & $d_{b^2-c^2}$           & 0.23      & 0.21      & 0.24 & 0.23                 \\ 
Cu-3\&4 & $d_{b^2-c^2}$        & 0.78      & 0.84      & 0.88 & 0.91                 \\ \hline
\end{tabular}
\label{Tab1:Occupation}
\end{table}

\textit{FS and nesting: a connection to high-$T_c$} -----
With the discussion above we demonstrated, that the bilayer structure of \bilayer\ satisfies one of the common ingredients for cuprates: Namely the low-energy physics of layer-2 is described by a single, almost half-filled orbital. However, contrary to the conventional CuO$_2$ planes, we have a CuO chain structure with a quasi 1D electronic structure, as can be seen from the FS in Fig.~\ref{fig:Fig5_DMFT_Fermi_Surface}(f). Already the parent compound \plane\ in Fig.~\ref{fig:Fig5_DMFT_Fermi_Surface}(c) has such a 1D character, but at quite a different filling. The Cu-1 FS of layer-1 in  Fig.~\ref{fig:Fig5_DMFT_Fermi_Surface}(d,e)
has instead a two-orbital character. However it differs from \original\ in Fig.~\ref{fig:Fig5_DMFT_Fermi_Surface}(a,b) --- not only by the filling (volume of the FS) but also because the Cu-2 sites are insulting. The latter cuts off, among others, the hopping of the Cu-1 $d_{z^2}$ orbital in the $y$ direction.

\begin{figure}[t]
\centering
\includegraphics[width=0.50\textwidth]{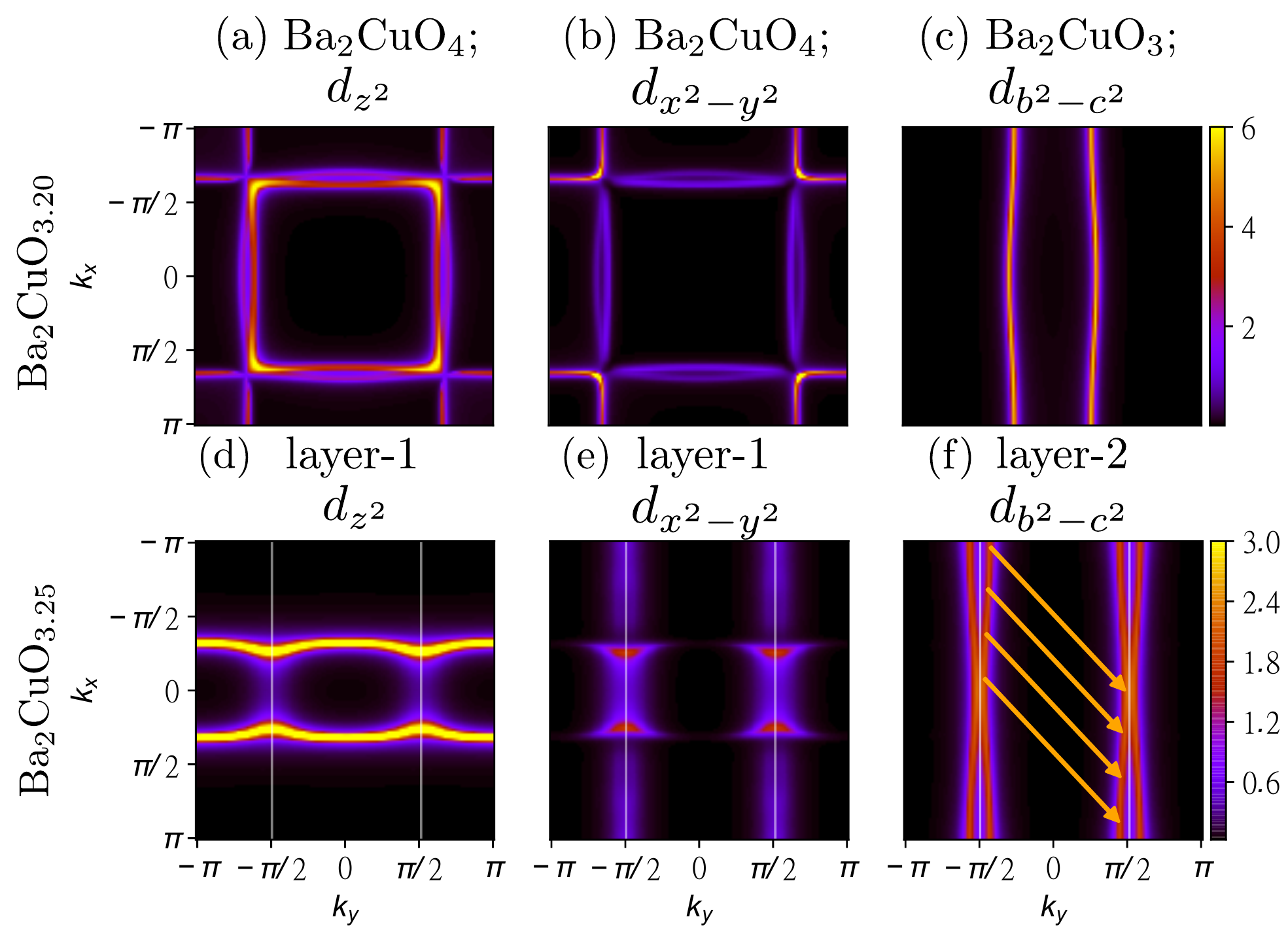}
\caption{DMFT FS at $k_z=0$ for the nominal doping of \experiment\ realized in the three crystal structures of Fig.~\ref{fig:Structure}: \original\ [(a) and (b) for $d_{z^2}$ and $d_{x^2-y^2}$], \plane\ (c), \bilayer\ layer-1 [(d) and (e) for $d_{z^2}$ and $d_{x^2-y^2}$ of Cu-1; Cu-2 is not shown because of its tiny occupation and insulating nature] and \bilayer\ layer-2 (f) where we also plotted the antiferromagnetic nesting vector.}
\label{fig:Fig5_DMFT_Fermi_Surface}
\end{figure}

Fig.~\ref{fig:Fig5_DMFT_Fermi_Surface}(f) shows that the correlation-induced charge transfer results not only in a $d_{b^2-c^2}$ orbital closer to half-filling but also in an almost perfectly nested FS. The nesting vector of $\bf{k_N} \simeq \rm{ \{\pi,\pi-\delta,0 \} }$, is similar to the dominant vector for commensurate, antiferromagnetic fluctuations ($\bf{k_{AF}} = \rm{ \{ \pi, \pi, 0 \} }$), and takes through its $k_x$ component also the slight warping of the FS into account.

\textit{Conclusion} -----
High temperature superconductivity remains one of the most puzzling phenomena in condensed matter physics and an overarching understanding is still missing. Hence, identifying common traits among superconductors, which help us to focus on the essential ingredients is of vital importance.
The recently discovered \experiment~superconductor challenges  the current picture of cuprate superconductivity. Its high hole  doping concentration,  compressed octahedra, and putative multi-band physics baffled the scientific community. In this work, we provide a resolution to the high doping of the compound by identifying a charge transfer process, which ultimately leads to one layer hosting a single-orbital   Fermi surface which is close to half-filling and almost perfectly nested. This bilayer structure also resolves the ambiguity, whether a \plane~or \original~structure is realized at the superconducting oxygen concentration of \experiment. Structural motives of both exist in the two layers, but the bilayer structure results in different dopings than previously thought.

Due to its strong nesting, layer-2 with its single orbital appears to be the natural candidate to host superconductivity. Layer-1, instead, serves more as a charge carrier reservoir/sink. This new insight shifts the focus from a highly overdoped to a quasi 1D superconductor, which is still in contrast to all other known cuprate superconductors that are 2D. Fluctuation exchange (FLEX) calculations
predict $p$-wave superconductivity or the pair density wave state for such an almost perfectly nested 1D system \footnote{This 1D chain is included in \cite{Shigeta2009,yoshida2021theory} as the extreme case of the anisotropic triangular lattice with hopping $t_2=0$.}.

Hopefully, future single crystals will allow for a better  crystallographic analysis, so that one can verify whether the bilayer structure is truly realized. Let us emphasize that the precise, in our case alternating, oxygen arrangement in layer-1 is of less relevance since this layer is not the one driving the system superconducting.
Even with the ideal, ordered oxygen arrangement Cu-1 and Cu-2  have a similar 3$d^{8.3}$ and 3$d^{8.2}$ \footnote{Note that,  on top of Table~\ref{Tab1:Occupation}, 2 electrons  fill the the $d_{a^2}$  of Cu-2 sites} electronic configuration, respectively, whereas the equivalent Cu-3$\&$4 sites in layer-2 have a distinct  3$d^{8.9}$ filling. This may explain the two-peak structure of the x-ray spectrum in \cite{Fumagalli2021}. Last but not least the prospects of a $p$-wave symmetry of the superconducting order parameter  at an unprecedented high $T_c$ calls for further experiments.

\emph{Note added.} When finalizing this work we became aware of a DFT electronic structure calculation
\cite{jin2021} for the bilayer structure  \bilayer.

\subsection{Acknowledgments}

\begin{acknowledgments}
We thank R.~Arita and A.~Prokofiev for helpful comments and discussions, as well as P. Kappl for proof-reading the manuscript. 
This work was supported by the  
Austrian Science Fund (FWF) through  projects P 30997, JSPS KAKENHI Grand Numbers JP20K22342 and JP21K13887. Calculations have been done on the Vienna Scientific Cluster (VSC).
\end{acknowledgments}

\end{document}